\documentclass[12pt]{article}

\newcommand{\ds}{\displaystyle}

\newcommand{\LI}{{\cal L}_{\rm int}^{(1)}}
\newcommand{\LII}{{\cal L}_{\rm int}^{(2)}}
\newcommand{\hI}{h_{\rm eff}^{(1)}}
\newcommand{\hII}{h_{\rm eff}^{(2)}}

\newcommand{\hef}[1]{h_{{\rm eff}#1}}

\newcommand{\psng}{\psi_{\rm NG}}
\newcommand{\pB}[1]{p_{\rm B}^{#1}}
\newcommand{\pF}[1]{p_{\rm F}^{#1}}

\newcommand{\chF}[1]{\chi_{{\rm F}#1}^{}}
\newcommand{\bchF}[1]{\bar{\chi}_{\rm F}^{\ #1}}
\newcommand{\bpsng}{\bar{\psi}_{\rm NG}}
\newcommand{\chng}{\chi_{\rm NG}}
\newcommand{\bchng}{\bar{\chi}_{\rm NG}}

\newcommand{\be}{\begin{equation}}
\newcommand{\ee}{\end{equation}}
\newcommand{\bea}{\begin{eqnarray}}
\newcommand{\eea}{\end{eqnarray}}

\newcommand{\hs}[1]{\hspace{#1mm}}
\newcommand{\vs}[1]{\vspace{#1mm}}

\newcommand{\g}[2]{g_{#1\bar{#2}}}
\newcommand{\gi}[2]{g^{#1\bar{#2}}}

\begin{document}

\begin{titlepage}
\null
\begin{flushright}
  {\tt hep-th/0204214}\\
TIT/HEP--477
\\
April  2002
\end{flushright}

\vskip 1cm
\begin{center}
{\LARGE \bf Low Energy Theorem for SUSY Breaking } 
{\LARGE \bf with 
Gauge Supermultiplets}
\lineskip .75em
\vskip 1.5cm

\normalsize

 {\large \bf 
Norisuke Sakai}
\footnote{{\it  e-mail address: nsakai@th.phys.titech.ac.jp}\\
Supported in part by Grant-in-Aid for Scientific 
Research from the Ministry of Education, Science and Culture 
No.13640269. 
},
~and~~ {\large \bf 
Ryo Sugisaka}
\footnote{{\it  e-mail address: sugisaka@th.phys.titech.ac.jp}\\
Supported 
by the Japan Society for the Promotion of Science for Young Scientists 
No.6665.
}

\vskip 0.5em

{ \it  
Department of Physics, Tokyo Institute of Technology\\
Tokyo 152-8551, JAPAN  }
\vspace{10mm}

{\bf Abstract}\\[5mm]
{\parbox{13cm}{\hspace{5mm}
%%%%%%%%%%%%%%%%%%%%%%%%%%%%%%%%%%%%%%%%%%%%%%%%%%%%%%%%%%%%
Low energy theorems of Nambu-Goldstone fermion 
associated with spontaneously broken supersymmetry 
are studied for gauge supermultiplets. 
Two possible terms in the effective Lagrangian are needed 
to deal with massless gaugino and/or massless gauge boson. 
As an illustrative example, a concrete model is worked out 
which can interpolate massless as well as massive gaugino 
and/or gauge boson to examine the low energy effective 
interaction of NG-fermion. 
%%%%%%%%%%%%%%%%%%%%%%%%%%%%%%%%%%%%%%%%%%%%%%%%%%%%%%%%%%%%
}}

\end{center}

\end{titlepage}

\renewcommand{\thefootnote}{\arabic{footnote}}
\baselineskip=0.7cm

\clearpage

%%%%%%%%%%%%%%%%%%%%%%%%%%%%%%%%%%%%%%
\section{
%{\large\bf
 Introduction}
%%%%%%%%%%%%%%%%%%%%%%%%%%%%%%%%%%%%%%%%%%%%%%%%%%%%%%%%%%

Supersymmetry (SUSY) is one of the most attractive ideas for unified 
model building \cite{DGSW}. 
When the SUSY is spontaneously broken, a massless particle 
 appears
 which is called the Nambu-Goldstone fermion
 \cite{FayetIl}, \cite{OR}. 
Its interaction is characterized by the broken SUSY 
and is typically summarized as low energy theorems
\cite{clark}--\cite{MSSS2}. 
The purpose of this paper is to study the low energy theorems 
for gauge supermultiplets, especially in the case of 
massless gaugino or gauge boson. 
We find that a new term should be used as the effective Lagrangian 
for massless gaugino. 
We also examine the effective interaction terms of NG fermion in an 
explicit model which can interpolate massless and massive 
particles in the gauge supermultiplet. 
This result should be useful even for supergravity theories 
provided the SUSY breaking scale is small enough, since gravitino 
behaves almost as an NG fermion for such a case \cite{Moroi}. 

If the supersymmetry (SUSY) is spontaneously broken, 
the boson-fermion mass-splitting is induced. 
There exists a Nambu-Goldstone (NG) fermion $\psi_{\rm NG}$ 
which shows up \cite{clark} 
in the supercurrent $J^{\mu}_{\alpha}(x)$ 
\be
 J^{\mu}_{\alpha}=\sqrt{2}if\left(\gamma^\mu \psng\right)_{\alpha}
 +J^{\mu}_{{\rm matter},\alpha}+\cdots, \label{super_cc_3D}
\ee
where $f$ is the order parameter of the SUSY breaking and 
$J^{\mu}_{{\rm matter},\alpha}(x)$ 
is the supercurrent for matter fields suitably dressed 
by the NG fermion. 
The low energy theorems have been worked out for chiral scalar 
supermultiplets. 
The effective coupling of the NG fermion is typically related to 
the mass-splitting of boson and fermion in the supermultiplet. 
In the case of gauge supermultiplet consisting of gauge boson $v_\mu$ and 
gaugino $\lambda$ for $U(1)$ gauge symmetry, 
the supercurrent 
$J^{\mu}_{{\rm matter},\alpha}(x)$ becomes
\begin{equation}
 J^{\mu}_{{\rm matter},\alpha} = 
 -i 
v_{\nu\rho}\left(\sigma^{\nu\rho}\sigma^{\mu}\right)_{\alpha\dot{\alpha}}
 \bar{\lambda}^{\dot{\alpha}} + \cdots ,
\end{equation}
where $v_{\mu \nu}$ is the gauge field strength. 
It has been shown that 
the single NG fermion interactions with a gauge supermultiplet 
can be expressed by the following effective Lagrangian 
\cite{lee-wu} for a massless gauge boson 
\be
{\cal L}_{\rm int} = \hef{} \psng\sigma^{\mu\nu}\lambda 
v_{\mu\nu} + {\rm h.c.}.\label{v:ND_mf=0}
\ee
Here the effective coupling constant $\hef{}$ is related to 
the gaugino mass $m_{\rm F}$ and the order parameter $f$ by
\be
\hef{} =\frac{m_{\rm F}}{\sqrt{2}f}.
\ee
This is the SUSY 
analog of the Goldberger-Treiman relation.

In Ref.\cite{MSSS2} we have extended this relation 
not only for a massless gauge boson but also for a massive one. 
In the case of $m_{{\rm F}^{}} \neq 0$ 
the effective coupling can be expressed 
in terms of mass-splitting of the gauge boson and the gaugino as 
\bea
\ds \LI \hs{-2}& = &\hs{-2} \ds \hI \psng \sigma^{\mu\nu}\lambda v_{\mu \nu} 
  + {\rm h.c.}+\cdots, \label{v:lag1}\\
\hI \hs{-2}&=&\hs{-2}-\frac{1}{\sqrt{2}f} 
\left( \frac{m_{\rm B}^2-m_{\rm F}^2}{m_{\rm F}} \right). \label{v:gtr}
\eea
Evidently this effective action is inadequate for massless gaugino 
$m_{\rm F}=0$. 
We will show that  the above effective Lagrangian (\ref{v:lag1})
does not contribute to the low energy theorem to the 
leading order of the Nambu-Goldstone fermion momentum. 
We will find that the following interaction term 
should be used instead as the effective Lagrangian 
for the case of 
massless gaugino $m_{{\rm F}^{}} = 0$ 
to describe the low energy theorem correctly 
\be
 \LII  = i \hII \psng  \sigma^\nu \bar{\lambda} v_\nu + {\rm h.c.} 
 .
 \label{v:lag2}
\ee
{}For massive gaugino and gauge boson, we will find that 
both terms are allowed and are equivalent. 
Therefore a linear combination of these terms is constrained 
by the low energy theorem 
\begin{equation}
\hII + \,m_{\rm F} \hI = - \frac{\Delta m^2}{\sqrt{2}f}, 
\qquad \Delta m^2 = m_{\rm B}^2-m_{\rm F}^2.
\end{equation}
{}For massless gauge boson $m_{\rm B}=0$, 
the new term (\ref{v:lag2}) is forbidden because of unbroken 
gauge invariance and one has to use the 
known effective Lagrangian (\ref{v:lag1}). 

In the next section, we work out the effective Lagrangian and low energy 
theorem. 
In sect.~3, we present a concrete model, which interpolate massive 
and massless gaugino and gauge boson, in order 
to examine the NG fermion interactions.

\section{
Low energy theorem for gauge supermultiplets}

We will consider interaction of NG fermion with momentum $q_{\rm NG}$ 
with gaugino of momentum $p_{\rm F}$ and 
gauge boson of momentum $p_{\rm B}$ and polarization 
$\epsilon(p_{\rm B})$ which satisfies 
$p_{\rm B}\cdot\epsilon(p_{\rm B})=0$. 
In the case of $m_{\rm B} \ne 0$ and $m_{\rm F} \ne 0$, 
these two Lagrangians can be regarded as equivalent 
in the soft NG fermion limit, $q_{\rm NG} \to 0$ as follows. 
Let us take matrix element of these Lagrangians 
between in-coming state of NG fermion and gaugino 
$|q_{\rm NG},\, \pF{}\rangle_{\rm in} $ and 
out-going state of gauge boson $|\pB{} \rangle_{\rm out} $. 
Using gaugino equation of motion, we obtain 
\bea
\hs{-10}
\ds \langle \pB{}|\LI 
 \hs{-3}&+&\hs{-3} \LII |q_{\rm NG};\pF{} \rangle \hs{50} 
 \nonumber \vs{2}   \\
  &=& \hs{-2} \ds
 i\hI \chi_{\rm NG}
   (\epsilon^{\ast}\hs{-1}\cdot \sigma)(\bar{\sigma}\cdot p_{{\rm B}}^{})\chF{}
 +i\hII  \chi_{\rm NG} 
 (\epsilon^{\ast}\hs{-1}\cdot\sigma) \bchF{} 
 \nonumber \\
 &=& \hs{-2} \ds
   i \left(\hI + \frac{\hII }{m_{\rm F}^{}} \right)
     \chi_{\rm NG}
   (\epsilon^{\ast}\hs{-1}\cdot \sigma)(\bar{\sigma}\cdot p_{{\rm B}}^{})\chF{}
    -q_{{\rm NG}\mu} i \left[ \frac{\hII }{m_{\rm F}}
     \chi_{\rm NG}
  ( \epsilon^{\ast}\hs{-1}\cdot\sigma) \bar{\sigma}^\mu \chF{} \right] 
   + {\rm h.c.} 
 \nonumber  \\ 
     \hs{3} &=& \hs{-2} \ds
    i \left( \hII + m_{\rm F}^{}\hI  \right)
       \chi_{\rm NG} (\epsilon^{\ast}\hs{-1}\cdot\sigma) \bchF{} +
      q_{{\rm NG}\mu} i \left[ \hI  \chi_{\rm NG}
  (\epsilon^{\ast} \cdot \sigma) \bar{\sigma}^\mu \chF{} \right]+ {\rm h.c.}. 
\label{matrix_L1L2}  
\eea
We see that these Lagrangians 
in Eqs.(\ref{v:lag1}) 
and (\ref{v:lag2}) 
contribute to the same matrix element 
ignoring higher $q^\mu_{\rm NG}$ terms. 
In this sense, we can use both interaction terms 
${\cal L}^{(1)}_{\rm int}$ and ${\cal L}^{(2)}_{\rm int}$ 
as the effective Lagrangians. 

To derive the relation between effective coupling constants 
$h^{(1)}_{\rm eff}, h^{(2)}_{\rm eff}$ and the gauge boson mass $m_{\rm B}$ 
and the gaugino mass $m_{\rm F}$, 
we introduce form factors $A_i(q^2), \: i=1, \cdots, 12$ 
 of the supercurrent $J^{\mu}_{\alpha}(x)$ between 
one-particle states of the 
gaugino $|\pF{}\rangle$ 
and the gauge boson $|\pB{}\rangle$ 
following our previous work \cite{MSSS2}
\begin{equation}
\begin{array}{l}
\langle \, \pB{}\, |\,J^{\mu}_\alpha(0)\,|\, \pF{}\, \rangle \vspace{3mm}
\\
 = \ \epsilon^{\ast}_\nu(\pB{}) \Bigl[
   A_1(q^2) q^\nu q^\mu 
 + A_2(q^2) q^\nu k^\mu 
 + A_3(q^2) q^\nu \sigma^\mu \bar{\sigma}^\rho q_\rho 
 + A_4(q^2) \eta^{\mu\nu} 
 + A_5(q^2)\sigma^\nu \bar{\sigma}^\mu
 \Bigr]_\alpha^{\ \ \beta} \chF{\beta}(\pF{}) \vspace{3mm}\\
 \ \ \ \ + \ \epsilon^\ast_\nu (\pB{}) \Bigl[
   A_6(q^2)q^\nu \sigma^\mu
 + A_7(q^2)q^\nu q^\mu \sigma^\rho q_\rho
 + A_8(q^2)q^\nu k^\mu \sigma^\rho q_\rho
 + A_9(q^2)\eta^{\mu \nu}\sigma^\rho q_\rho  \vspace{3mm} 
\\
 \hspace{40mm}
+ A_{10}(q^2)q^\mu \sigma^\nu
 + A_{11}(q^2)\sigma^\nu \bar{\sigma}^\rho \sigma^\mu 
k_\rho
 + A_{12}(q^2)\sigma^\mu \bar{\sigma}^\rho \sigma^\nu 
q_\rho
 \Bigr]_{\alpha \dot{\beta}}\bchF{\dot{\beta}}(\pF{}), 
\end{array} \label{v:matrix_J}
\end{equation}
where $q^\mu = \pB{\mu}-\pF{\mu}$, $k^\mu = \pB{\mu } + \pF{\mu }$. 
The supercurrent conservation $\partial_\mu J^\mu_\alpha=0$ 
gives a relation among the form factors in terms of mass-splitting 
$\Delta m^2 \equiv m_{\rm B}^2 - m_{\rm F}^2$ 
between the gauge boson and the gaugino
\begin{equation}
 q^2 \left[ A_{10}(q^2) + A_{11}(q^2)-A_{12}(q^2) \right]
  = -2 \Delta m^2 A_{11}(q^2). 
  \label{current_conservation}
\end{equation}

As is usual in current algebra, the supercurrent matrix element 
can have a massless 
NG fermion pole. 
So some of form factors in Eq.(\ref{v:matrix_J}) are singular 
in the limit of $q^2 \to 0$. 
Eq.(\ref{super_cc_3D}) implies a 
non-vanishing matrix element of the supercurrent 
between the vacuum and the single NG fermion state 
$|\,q,\,\psng \rangle$ with momentum $q^\mu$ 
\be
\langle\,0\,|\,J_\alpha^\mu(0)\,|\,q,\,\psng \rangle = 
\sqrt{2}if\sigma^{\mu}_{\alpha\dot{\alpha}}\bar{\chi}_{\rm NG}^{\dot{\alpha}},
\ee
where $f$ is the order parameter of the SUSY breaking. 
Since the 
combination $J^{\mu}_{\alpha}-\sqrt{2}
i f \sigma^{\mu}_{\alpha\dot{\alpha}}\bar{\psi}_{\rm NG}^{\dot{\alpha}}$ 
has vanishing matrix element between the vacuum and the single NG fermion 
state, 
its matrix element 
between gauge boson and gaugino states 
becomes non-singular in the limit $q^2 \to 0$. 
Now we define an NG fermion source 
$j^{\rm NG}_{\alpha}(x)$ 
by using the NG fermion field $\psng(x)$ 
\be
 j^{\rm NG}_{\alpha}(x)\equiv -i\sigma^{\mu}_{\alpha\dot{\alpha}}
 \partial_{\mu} \bar{\psi}_{\rm NG}^{\dot{\alpha}}(x). \label{def_NGs}
\ee
By introducing form factors $B_i(q^2), \: i=1, \cdots, 4$ 
of the NG fermion source, we obtain 
\begin{eqnarray}
 \langle \,\pB{}\, |\,j_{\alpha}^{\rm NG}(0)|\, \pF{}\, \rangle \ = \ 
 \epsilon^{\ast}_\nu(\pB{}) \left[
   B_{1}(q^{2}) q^{\nu} 
 + B_{2}(q^{2}) q_{\rho}\sigma^{\rho}\bar{\sigma}^{\nu}
 \right]_{\alpha}^{\ \ \beta} \chF{\beta}(\pF{})
  \vspace{1mm} \nonumber\\
\hspace{5cm} + \ \epsilon^{\ast}_{\nu}(\pB{}) \left[
   B_{3}(q^{2})q^{\nu}\sigma^{\rho}q_{\rho} 
 + B_{4}(q^{2})\sigma^{\nu}
 \right]_{\alpha \dot{\beta}}\bchF{\dot{\beta}}(\pF{}),
 \label{v:matrix_j_NG}
\end{eqnarray}
\begin{eqnarray}
 \langle\,\pB{}\, |\,\bar{\psi}_{\rm NG}^{\dot{\alpha}}(0)\,|\, \pF{}\,
  \rangle  \ = \ \epsilon^{\ast}_\nu(\pB{}) \left[
 - \displaystyle{\frac{B_{1}(q^{2})}{q^{2}}} 
q^{\nu}\bar{\sigma}^{\rho} q_\rho
 + B_{2}(q^{2})\bar{\sigma}^{\nu}
 \right]^{\dot{\alpha}\beta} \chF{\beta}(\pF{})
 \vspace{1mm} \nonumber \\
\hspace{5cm} + \ \epsilon^{\ast}_{\nu}(\pB{}) \left[
 B_{3}(q^{2})q^{\nu} 
 - \displaystyle{\frac{B_{4}(q^{2})}{q^{2}}}q_{\rho} 
\bar{\sigma}^{\rho}
 \sigma^{\nu}  \right]^{\dot{\alpha}}_{\ \ \dot{\beta}}
 \bchF{\dot{\beta}}(\pF{}).\label{matrix_lambda}
\end{eqnarray}
Form factors in the combination 
$J^\mu_\alpha -\sqrt{2}if \sigma^{\mu}_{\alpha \dot{\alpha}} 
\bar{\psi}_{\rm NG}^{\dot{\alpha}}$ 
must be regular in the limit $q^2 \rightarrow 0$ 
\begin{equation}
\begin{array}{l}
\langle \, \pB{}\, |\,J^{\mu}_\alpha(0)-\sqrt{2}if 
\sigma^{\mu}_{\alpha \dot{\alpha}} 
\bar{\psi}_{\rm NG}^{\dot{\alpha}}\,|\, \pF{}\, \rangle \vspace{1.5mm}
\\
 \ \ \ = \ \epsilon^{\ast}_\nu(\pB{}) \Biggl[
   \ds A_1(q^2) q^\nu q^\mu 
 + A_2(q^2) q^\nu k^\mu 
 + \left( A_3(q^2) +\sqrt{2} if \frac{B_1(q^2)}{q^2}\right)q^\nu 
 \sigma^\mu \bar{\sigma}^\rho q_\rho \Bigr. \vspace{1.5mm} \\
 \Bigl.\hs{20}+ \left(A_4(q^2) + 2\sqrt{2}ifB_2(q^2)\right) \eta^{\mu\nu}  
 + \left(A_5(q^2) + \sqrt{2}ifB_2(q^2)\right)\sigma^\nu \bar{\sigma}^\mu
 \Biggr]_\alpha^{\ \ \beta} \chF{\beta}(\pF{}) \vspace{2mm}\\
  \ \ \ \ + \ \epsilon^\ast_\nu (\pB{}) \Biggl[
   \left( A_6(q^2) -\sqrt{2}ifB_3(q^2)\right) q^\nu \sigma^\mu
 + A_7(q^2)q^\nu q^\mu \sigma^\rho q_\rho
 + A_8(q^2)q^\nu k^\mu \sigma^\rho q_\rho
 \vspace{2mm}
\\
 \hspace{55mm}
 + A_9(q^2)\eta^{\mu \nu}\sigma^\rho q_\rho + A_{10}(q^2)q^\mu \sigma^\nu
 + A_{11}(q^2)\sigma^\nu \bar{\sigma}^\rho \sigma^\mu 
k_\rho \vspace{2mm} \\
 \ds \hs{72}+ \left( A_{12}(q^2) +\sqrt{2}if\frac{B_4(q^2)}{q^2} \right)
  \sigma^\mu \bar{\sigma}^\rho \sigma^\nu 
q_\rho
 \Biggr]_{\alpha \dot{\beta}}\bchF{\dot{\beta}}(\pF{}). 
\end{array} \label{v:matrix_J-NG}
\end{equation}
Therefore we obtain 
 the singularity of the form factors 
$A_{3}(q^2)$ and $A_{12}(q^2)$ at 
$q^2 = 0$ unless $B_1(0)$ and $B_4(0)$ vanish respectively
\bea
 \lim_{q^2 \to 0} q^2A_{3}(q^2) = -\sqrt{2}if B_1(0), 
 \qquad 
 \lim_{q^2 \to 0} q^2A_{12}(q^2) = -\sqrt{2}if B_4(0). \label{A12-B4}
\eea
Combining Eq.(\ref{A12-B4}) with Eq.(\ref{current_conservation}) 
in the limit $q^2 \to 0 $ gives 
\begin{equation}
 \sqrt{2}if B_4(0) = -2 \Delta m^2 A_{11}(0).  \label{B4_A11}
\end{equation}

To relate the form factor $B_4(0)$ to an effective coupling constant
 of the NG fermion with the gauge boson and the gaugino, 
we evaluate a transition amplitude between the in-state 
$|q;\pF{}\rangle_{\rm in}$ 
and the out-state $|\pB{}\rangle_{\rm out}$.
This S-matrix element can be expressed by using an effective 
interaction Lagrangian 
${\cal L}_{\rm int}$ as
\bea
 \mbox{}_{\rm out}\langle \,\pB{}|\,q;\,\pF{}\rangle_{\rm in}&\!\!=&\!\!
  \mbox{}_{\rm I}\langle \pB{} |e^{iS_{\rm int}(x)}
  |q;\pF{}\rangle_{\rm I}  \nonumber\\
 &\simeq & \hs{-2} i \int\! {\rm d}^4x \mbox{}_{\rm I}\langle \,\pB{}| 
 e^{-iPx}{\cal L}_{\rm int}(0)
  e^{iPx}|q;\pF{}\rangle_{\rm I} \nonumber \\
 &= & \hs{-2} i(2\pi)^4\delta^4(\pB{}-\pF{}-q)\:
 \mbox{}_{\rm I}\langle \pB{}|{\cal L}_{\rm int}(0)|q;\pF{}\rangle_{\rm I}, 
 \label{B-NG_F}
\eea
where $|\pB{}\rangle_{\rm I}$ and $|q;\pF{}\rangle_{\rm I}$  
denote 
states in the interaction picture.
On the other hand, using the LSZ reduction formula in four dimensions, 
it can also be written as 
\bea
\mbox{}_{\rm out}\langle \,\pB{}|\,q;\,\pF{}\rangle_{\rm in} \hs{-2}
 &=& \hs{-2}-i \int \! {\rm d}^4 x\, e^{iqx} \bar{v}_{\rm NG}^{}
  i \gamma^\mu \partial_\mu 
 \mbox{}_{\rm I}\langle \pB{} |\Psi_{\rm NG}(x) |\pF{}\rangle_{\rm I}
   \nonumber \\
&=& \hs{-2} -i(2\pi)^4 \delta^4 (\pB{}-\pF{}-q) \chi_{\rm NG}^{}(q)
 q_{\mu} \sigma^{\mu}
 \mbox{}_{\rm I} \langle \pB{} |\bpsng(0)|\pF{} \rangle_{\rm I} 
\nonumber\\
 &\!\!\!&\!\!-i(2\pi)^4 \delta^4 (\pB{}-\pF{}-q) \bchng(q) q_{\mu}
 \bar{\sigma}^{\mu}
 \mbox{}_{\rm I}\langle\pB{} |\psng(0)|\pF{}\rangle_{\rm I}, \label{B-NG_F2}
\eea
where Dirac spinors $\Psi_{\rm NG}$ and $\bar{v}_{\rm NG}$
 are decomposed into Weyl spinors as
\be
\Psi_{\rm NG} \equiv \left( \hs{-1.5}
\begin{array}{l}
\psng \\ \bar{\psi}_{\rm NG}^{} 
\end{array} \hs{-1.5} \right) , \hs{5}
\bar{v}_{\rm NG} \equiv \left( \chng, \,\bchng \right).
\ee
Since we do not need to distinguish the interaction picture 
and the 
Heisenberg picture for one-particle states, we drop the 
subscript I 
for one-particle states in the following. 
Therefore Eqs.(\ref{B-NG_F}) and (\ref{B-NG_F2}) 
lead to a relation between matrix elements of 
the interaction Lagrangian and NG fermion source 
\bea
 \langle\pB{} |{\cal L}_{\rm int}(0)|q;\pF{}\rangle_{\rm I} \hs{-2}&=&\hs{-2} 
 \ds -\chng(q)q \cdot \sigma
  \langle\pB{} |\bpsng(0)|\pF{}\rangle
 \ds -\bchng(q)q \cdot \bar{\sigma}
 \langle\pB{} |\psng(0)|\pF{}\rangle \nonumber \vs{1.5}\\
 \hs{-2}&=&\hs{-2} 
 \ds -\chng(q) \langle\pB{} |j^{\rm NG}(0)|\pF{}\rangle
 -\bchng(q)   \langle\pB{} |\bar{\jmath}^{\rm NG}(0)|\pF{}\rangle.
 \label{ME-of-Lint}
\eea
Comparing Eqs.(\ref{matrix_L1L2}) and (\ref{ME-of-Lint}) 
with Eq.(\ref{v:matrix_j_NG}), we obtain  
\begin{eqnarray}
&\hs{-3}\chng(q) \epsilon^{\ast}_\nu(\pB{}) \left[
  \ds \left\{ B_{1}(0)-2B_{2}(0) \right\} q^{\nu} 
  + \left\{ B_{2}(0) +i\hI \right\} \sigma^\nu \bar{\sigma} 
  \cdot q \,\right]^{\dot{\alpha}\beta} 
  \chF{\beta}(\pF{})
 \vspace{1mm} \nonumber \\
&\hs{10} \ds + \ \chng(q) \epsilon^{\ast}_{\nu}(\pB{}) \left[
   B_{3}(0)q^{\nu} q \cdot \sigma 
 + \{ B_4(0) + i \hII \} \sigma^\nu 
 \right]^{\dot{\alpha}}_{\ \ \dot{\beta}} 
 \bchF{\dot{\beta}}(\pF{})= 0. \label{B-g}
\end{eqnarray}
For the case of $ m_{\rm F}^{} \neq 0 $, 
we can use the gaugino equation of motion
\be
\bchF{} = \frac{1}{m_{\rm F}^{}} \bar{\sigma} \cdot \pF{} \,\chF{} 
= \frac{1}{2m_{\rm F}^{}}  \bar{\sigma} \cdot (k-q) \, \chF{},
\label{v:eom_lm}
\ee
which gives relations among the form factors 
and the couplings $\hI,\hII$ 
\be
 B_4(0) = -i \left(\hII + m_{\rm F} \hI\right), 
 \hs{5}\frac{1}{2}B_1(0) = B_2(0) = -\frac{B_4(0)+i \hII}{m_{\rm F}},
 \hs{5}  B_3(0) = 0.
 \label{relation_B1-B4_1}
\ee
Thus Eq.(\ref{A12-B4}) becomes
\begin{equation}
 -\sqrt{2}(\hII + m_{\rm F}\hI) f =  2\Delta m^2
A_{11}(0).  \label{v:GTR}
\end{equation}

Let us consider the case of $m_{\rm F}=0$. 
 Eq.(\ref{matrix_L1L2}) shows 
that the Lagrangian $\LI$ does not contribute to 
the matrix element in leading orders of $q_{\rm NG}^\mu$ 
in this case. 
{}For $m_{\rm F}=0$, Eq.(\ref{B-g}) leads to
\be
 B_4(0) = -i \hII, \hs{5}\frac{1}{2}B_1(0) 
 = B_2(0) = -i\hI, \hs{5} B_3(0) = 0.
 \label{relation_B1-B4}
\ee
Therefore 
the effective Lagrangian consists solely 
of ${\cal L}^{(2)}_{\rm int}$ and its coupling $h^{(2)}_{\rm eff}$ is 
given in terms of the mass-splitting $\Delta m^2$ 
for the case of massless gaugino $m_{\rm F} =0$ 
\begin{equation}
 -\sqrt{2}\hII f =  2\Delta m^2
A_{11}(0).  \label{v:GTR2}
\end{equation}

Now we consider the case of massless gauge boson $m_{\rm B} =0$. 
In this case, we have an unbroken gauge symmetry, 
which requires the vanishing matrix element of Eq.(\ref{ME-of-Lint}), 
if we replace $\epsilon^\ast_\mu(\pB{})$ 
by $p_{{\rm B}\mu}$. 
Using  Eq.(\ref{v:matrix_j_NG}), we obtain relations among the 
form factors $B_i(0)$ 
\be
\frac{1}{2}B_1(0) = B_2(0) = -\frac{B_4(0)}{m_{\rm F}},\hs{5}  B_3(0) = 0.
\label{mB=0_B1-B4}
\ee
These relations combined with Eq.(\ref{relation_B1-B4}) imply 
$\hII =0$. 
 Therefore gauge invariance forbids $\LII$ 
as a piece of effective Lagrangian for $m_{\rm B}=0$. 

We can determine the form factor $A_{11}(0)$ 
by substituting the following expression of the supercurrent 
into Eq.(\ref{v:matrix_J-NG}) in the limit $q^2\to 0$
\begin{equation}
 J^\mu_\alpha -\sqrt{2}if \sigma^{\mu}_{\alpha \dot{\alpha}}
 \bar{\psi}_{\rm NG}^{\dot{\alpha}} = 
 -i 
v_{\nu\rho}\left(\sigma^{\nu\rho}\sigma^{\mu}\right)_{\alpha\dot{\alpha}}
 \bar{\lambda}^{\dot{\alpha}} + \cdots \ ,
\end{equation}
where $\cdots$ denotes higher order terms with NG fermion 
and possible interaction terms. 
Neglecting 
possible ``renormalization effects'' due to 
interactions and higher order terms, 
we find 
\begin{equation}
 A_{11}(0)= \frac{1}{2},
\end{equation}
which leads to the SUSY Goldberger-Treiman relation 
\be
\ds \hII + \,m_{{\rm F}} \hI = - \frac{\Delta m^2}{\sqrt{2}f}. \label{GTR}
\ee

Now we consider the case where fermion mass-eigenstates 
$\{\tilde{\lambda}_i\}$ are mixtures of gauginos $\{\lambda_i\}$ 
which are superpartners of gauge bosons $\{(v_i)_{\mu}\}$ as 
\be
\lambda_i(x) = V_{i,j} \tilde{\lambda}_j(x),
\label{v:mix_lm}
\ee
where $V_{i,j}$ is the unitary mixing matrix.
In this case, the effective Lagrangian should be written 
in terms of mass-eigenstates as follows
\be
\ds {\cal L}_{\rm int} = \sum_{i,j}\hef{i,j}^{(1)} 
\psng \sigma^{\mu\nu}\tilde{\lambda}_j (v_i)_{\mu \nu} 
+ \sum_{i,j}\hef{i,j}^{(2)} i \psng  \sigma^\nu 
\bar{\tilde{\lambda}}_j (v_i)_\nu + {\rm h.c.}, 
\ee
and thus Eq.(\ref{v:GTR}) becomes
\be
 \ds -\sqrt{2}(\hef{i,j}^{(2)} + \,m_{{\rm F},j} \hef{i,j}^{(1)}) f 
 = 2\Delta m^2_{i,j} A_{11}(0)_{i,j},  \label{v:GTR_mix}
\ee
where $m^2_{i,j} = m_{{\rm B},i}^2 - m_{{\rm F},j}^2$.
Taking account of the mixing Eq.(\ref{v:mix_lm}), the supercurrent becomes
\be
 J^\mu_\alpha = \sqrt{2}if \sigma^{\mu}_{\alpha \dot{\alpha}}
 \bar{\psi}_{\rm NG}^{\dot{\alpha}} 
 -i \sum_{k, l} V_{k,l} \,(v_k)_{\nu\rho}\left(\sigma^{\nu\rho}
 \sigma^{\mu}\right)_{\alpha\dot{\alpha}} 
 \bar{\tilde{\lambda}}_l^{\dot{\alpha}} + \cdots \ ,
\ee
which determines the value of the form factor $A_{11}(0)_{i,j}$ as
\be
A_{11}(0)_{i,j}=\frac{1}{2}V_{i,j},
\ee
and so Eq.(\ref{GTR}) can be written as
\be
\ds \hef{i,j}^{(2)} + \,m_{{\rm F},j} \hef{i,j}^{(1)} 
= - \frac{\Delta m^2_{i,j}}{\sqrt{2}f}V_{i,j}.
\label{eq:mixing}
\ee

In summary, we should use 
the following effective Lagrangians 
\bea
m_{\rm F} \ne 0,m_{\rm B} =0 :& \ds {\cal L}_{\rm int} 
\hs{-2}&= \hI \psng \sigma^{\mu\nu}\lambda v_{\mu \nu}  + {\rm h.c.}, \\
m_{\rm F}  = 0 ,m_{\rm B} \ne 0 :&  {\cal L}_{\rm int}  
\hs{-2}&= i \hII \psng  \sigma^\nu \bar{\lambda} v_\nu + {\rm h.c.}, \\
m_{\rm F} \ne 0,m_{\rm B} \ne0 :& \ds {\cal L}_{\rm int} 
\hs{-2}&= \hI \psng \sigma^{\mu\nu}\lambda v_{\mu \nu} 
  +   i \hII \psng  \sigma^\nu \bar{\lambda} v_\nu + {\rm h.c.}.
\eea
The corresponding 
effective couplings 
are given by the 
SUSY Goldberger-Treiman relations 
\bea
m_{\rm F} \ne 0,m_{\rm B} =0 :& \ds \hI \hs{-2} 
&= \frac{m_{\rm F}}{\sqrt{2}f},
\label{v:GTR_1} \\
m_{\rm F}  = 0 ,m_{\rm B} \ne 0 :& \ds \hII \hs{-2}&=
 - \frac{m_{\rm B}^2}{\sqrt{2}f},
\label{v:GTR_2} \\
m_{\rm F} \ne 0,m_{\rm B} \ne0 :& \ds \hII \hs{-3}&
+ \,m_{\rm F} \hI = - \frac{\Delta m^2}{\sqrt{2}f}, 
\qquad \Delta m^2 = m_{\rm B}^2-m_{\rm F}^2.
\label{v:GTR_12}
\eea
If there is a mixing for gaugino, 
we only need to multiply them by the mixing matrix $V_{i,j}$ 
as in Eq.(\ref{eq:mixing}).

\section{
Low energy theorem 
in a concrete model}

So far 
we derived the SUSY Goldberger-Treiman relation 
for gauge supermultiplet not only with massless gauge boson as 
in Ref.\cite{lee-wu}
 but also with massive one. 
 We will examine and check 
 the result in Eqs.(\ref{v:GTR_1}) - (\ref{v:GTR_12}) 
 by a spontaneously broken SUSY model interpolating 
 $m_{\rm F}=0$ and $m_{\rm F}\not=0$.

The model considered here is the spontaneously broken 
$U(1)$ gauge theory  in four dimensions 
which gives mass for the gauge boson and the gaugino. 
In order to have NG fermion without containing gaugino component, 
 SUSY is broken by the O'Raifeartaigh mechanism\cite{OR}.  
We introduce the following Lagrangian with  chiral superfields 
$\Phi^{(i)}=(A^{(i)},\psi^{(i)},F^{(i)}),\,i=0,1,2$ 
 neutral under $U(1)$ group, and chiral superfields 
$\Phi^+,\Phi^-$ with 
$U(1)$ charge $\pm e$, respectively 
\bea
{\cal L} \hs{-2}&=&\hs{-2} \ds\left[\, \sum_{i=0}^2 
\bar{\Phi}^{(i)}\Phi^{(i)} + \bar{\Phi}^+ e^{eV} \Phi^+ 
+ \bar{\Phi}^- e^{-eV} \Phi^- + \alpha \bar{\Phi}^{(0)}\Phi^+\Phi^- 
+ \alpha {\Phi}^{(0)} \bar{\Phi}^+ 
\bar{\Phi}^-\,\right]_{\theta^{2}\bar{\theta}^{2}} \nonumber \\
 \hs{-2}&&\hs{-2} \ds + \left[ \, \frac{1}{4} W^\alpha W_\alpha 
 + \frac{\Phi}{M} W^\alpha W_\alpha 
 + P(\Phi^{(i)},\Phi^\pm)  \right]_{\theta^{2}} 
 + {\rm h.c.}. \label{lag}
\eea
The gauge kinetic function has a non-minimal piece with the parameter 
 $1/M$ and is chosen to be proportional to the NG fermion superfield 
 $\Phi = (A,\psi,F), \;$  $\psi = \psng,$ $\langle F\,\rangle = -f$, 
 which is determined later. 
 This non-minimal kinetic term is added to give mass term 
 for gaugino bilinear. 
The K$\ddot{\rm a}$hler potential has a 
non-minimal piece with the 
parameter $\alpha$ of mass dimension $-1$. 
This term 
is added to allow a possibility of massless fermion containing 
gaugino component as we show below. 
The superpotential $P$ is given in terms of 
parameters $\ell$ of mass dimension two, 
$m, k$ and $h$ of mass dimension one 
and dimensionless $g$ 
\bea
\ds P(\Phi^{(i)},\Phi^\pm)  
= \ell \Phi^{(2)} + m \Phi^{(0)}\Phi^{(1)} 
+ g\Phi^{(1)}\Phi^{(1)}\Phi^{(2)} - \frac{k^2}{h}\Phi^+\Phi^- 
+ \frac{1}{2h}(\Phi^+\Phi^-)^2.
\label{SP}
\eea

Now we consider the minimum of the potential for scalar fields 
\be
{\cal V} = \frac{1}{2} D^2 + \sum_{i, \bar \jmath} \gi{i}{\jmath} 
\frac{\partial P}{\partial A^i} 
\frac{\partial \bar P}{\partial \bar{A^{j}} },
\ee
where $\g{i}{\jmath}$ is K$\ddot{\rm a}$hler metric in field space. 
{}For the case of $g\ell < 0$ and $2|g\ell| \ge m^2/(1-\beta^2)$ 
where $\beta \equiv \sqrt{2}\alpha k$, 
we obtain an absolute minimum with a relation 
\begin{equation} 
\langle A^{(2)} \rangle = -\frac{m}{2g\langle A^{(1)} \rangle} 
 \langle A^{(0)} \rangle, 
 \qquad 
 \langle A^{+} \rangle = \langle A^{-} \rangle
\end{equation}
In order to simplify the determination of the explicit value of the fields, 
we choose to impose the following fine tuning of parameters of the 
model 
\be
h = \frac{\alpha mv}{2(1-\beta^2)k^2}, 
\qquad 
 v \equiv \langle A^{(1)} \rangle .
\ee
We obtain 
the minimum of the scalar potential 
at the following value for 
fields 
\bea
 \ds \langle A^{(1)} \rangle \hs{-2}
 &
 = 
 &
 \sqrt{\frac{2|g\ell|-m^2/(1-\beta^2)}{2g^2}},
 \qquad 
\langle A^{+} \rangle %\hs{-2}
 =
 \langle A^{-} \rangle = k,
\eea
We obtain a flat direction along $\langle A^{(0)} \rangle $ 
and choose $\langle A^{(0)} \rangle =0$ in the following. 
In this vacuum, both SUSY and $U(1)$ gauge symmetry are broken. 
So the gauge boson becomes massive $m_{\rm B}=ek$, 
 and there is non-zero vacuum energy 
caused by $\langle F^{(0)} \rangle, \langle F^{(2)} \rangle, 
\langle F^\pm \rangle \ne 0$.

To find an NG fermion, we focus on fermion mass terms. 
Since $\tau \equiv (\psi^+-\psi^-)/i\sqrt2$ and gaugino 
$\lambda$ mixes together and do not contain NG fermion component, 
we consider first 
possible massless fermions arising 
from a mixing of $\xi \equiv (\psi^+ + \psi^-)/\sqrt{2}$, and 
$\psi^{(0)}, \psi^{(1)}, \psi^{(2)}$. 
Because of non-minimal K$\ddot{\rm a}$hler potential, 
the kinetic terms of the fields $\Phi^{(1)}, 
\Phi^+$ and $\Phi^-$ become non-canonical. 
We define the following fermions 
to normalize the kinetic terms
\be
\begin{array}{c}
\zeta \equiv \sqrt{\frac{1+\beta}{2}} 
\left( \psi^{(0)} + \xi \right) \vs{2}\\ 
\eta \equiv \sqrt{\frac{1-\beta}{2}} 
\left( \psi^{(0)} - \xi \right)
\end{array}
\hs{5}
\Leftrightarrow
\hs{5}
\begin{array}{c}
\psi^{(0)}= \frac{1}{\sqrt{2(1+\beta)}} \zeta 
+ \frac{1}{\sqrt{2(1-\beta)}} \eta \vs{1.5}\\
\xi =  \frac{1}{\sqrt{2(1+\beta)}} \zeta 
- \frac{1}{\sqrt{2(1-\beta)}} \eta
\end{array}
\begin{array}{l}
  \\ \hs{-2}. 
\end{array}
\label{eq:mixing_massless_fermion}
\ee
The mass matrix of these canonically normalized fermions 
becomes
\be
-\frac{1}{2}(\psi^{(1)}, \psi^{(2)}, \zeta\,,\, \eta\,\,)\,\left(
\begin{array}{cccc}
0 & 2gv & \frac{m}{\sqrt{2(1+\beta)}} &\frac{m}{\sqrt{2(1-\beta)}}\\
2gv & 0 & 0 & 0 \\
\frac{m}{\sqrt{2(1+\beta)}} & 0 & 0 & 0 \\
\frac{m}{\sqrt{2(1-\beta)}} & 0 & 0 & 0 
\end{array} \right) \left(
\begin{array}{c}
\psi^{(1)} \\ \psi^{(2)}\\ \zeta \\ \eta 
\end{array} \right)
\begin{array}{l}
  \\  \\ \\ \hs{-2}.
\end{array}
\ee
We find two zero-eigenvalues for the mass matrix 
and the corresponding zero-eigenmodes as 
\bea
\tilde{\psi}_{01}\hs{-2}& =&\hs{-2} 
\ds\frac{1}{\sqrt{m^2 + 4g^2v^2(1-\beta^2)}} \left( -m \psi^{(2)} 
+ 2gv (1-\beta) \sqrt{\frac{1+\beta}{2}} \zeta  
+  2gv(1+\beta)\sqrt{\frac{1-\beta}{2}} \eta \,\right), \nonumber\\
\tilde{\psi}_{02} \hs{-2}& =&\hs{-2} \sqrt{\frac{1+\beta}{2}} \zeta  
- \sqrt{\frac{1-\beta}{2}} \eta \ .
\label{zero-modes}
\eea

To identify the NG fermion from a linear 
combination of these zero-modes, 
we consider their SUSY transformation 
\bea
\delta_\epsilon \tilde{\psi}_{01} \hs{-2}
& =&\hs{-2} -\sqrt{2} f \epsilon , \label{NG_SUSY_tr2}\\
\delta_\epsilon \tilde{\psi}_{02} \hs{-2}& =&\hs{-2} 0 ,
\eea
where $f$ is the order parameter of SUSY breaking 
and is given by the square root of the vacuum energy
\be
f^2 = \gi{0}{0} \frac{\partial P}{\partial A^{(0)}} 
\frac{\partial \bar P}{\partial \bar{A}^{(0)}} 
+ \frac{\partial P}{\partial A^{(2)}} 
\frac{\partial \bar P}{\partial \bar{A}^{(2)}} 
= \frac{m^2}{4g^2(1-\beta^2)^2} \{m^2+4g^2v^2(1-\beta^2)\}.
\ee
Since one of zero-modes $\tilde{\psi}_{02}$ has 
no contribution to SUSY breaking, 
 we identify $\tilde{\psi}_{01}$ 
as the NG fermion $\psng$ which can be rewritten using 
(\ref{eq:mixing_massless_fermion}) as
\be
\psng = \frac{1}{\sqrt{m^2 + 4g^2v^2(1-\beta^2)}} 
\left( 2gv (1-\beta^2) \psi^{(0)} -m \psi^{(2)}\right).
\ee
Thus the NG fermion superfield $\Phi$ in our model (\ref{lag}) 
can be chosen as
\be
\Phi = \frac{1}{\sqrt{m^2 + 4g^2v^2(1-\beta^2)}} 
\left( 2gv (1-\beta^2) \Phi^{(0)} -m \Phi^{(2)}\right).
\ee
In this case, Lagrangian (\ref{lag}) has 
the following fermion mass terms 
between $\tau  \equiv (\psi_+-\psi_-)/i\sqrt{2}$ and 
gaugino $\lambda$ 
\be
{\cal L}_{{\rm mass}} = -\frac{1}{2} \left( -2k^2h \right) 
\tau^\alpha \tau_\alpha -\frac{1}{2} 
\left( \frac{2\langle F \rangle}{M} \right)
\lambda^\alpha \lambda_\alpha - (ek) 
\tau^\alpha \lambda_\alpha + {\rm h.c.},
\ee
where $\langle F\,\rangle = -f$ and the NG fermion interaction terms
\be
{\cal L}_{\rm int}= i \frac{e}{2}
\frac{-2gv\beta}{\sqrt{m^2 + 4g^2v^2(1-\beta^2)}}\psng\, 
\sigma^\mu \, \bar \tau\,v_\mu 
- \frac{\sqrt{2}}{M} \psi_{\rm NG}
\,\sigma^{\mu\nu}\lambda v_{\mu\nu} + {\rm h.c.}.
\label{int_lag}
\ee
The mass matrix between $\lambda$ and $\tau$ becomes
\be
-\frac{1}{2}\hspace{-1.5mm}
 \begin{array}{cc} 
  \bigl( \tau^\alpha &\hs{-3}\lambda^\alpha \bigr)
 \end{array}
 \hspace{-2mm} \left( \hspace{-1mm} 
 \begin{array}{cc}
  a &\hspace{-1mm} m_{{\rm B},2}^{} \\
  m_{{\rm B},2}^{} & \hspace{-2mm} b
 \end{array}
\hspace{-1mm} \right)\hspace{-1mm} \left( \hspace{-2mm} 
 \begin{array}{c} 
  \tau_{\alpha}\\
  \lambda_{\alpha}
 \end{array}
\hspace{-2mm} \right),  \hs{5} m_{{\rm B},2}^{} \equiv ek, 
\hs{5} a \equiv -2k^2h,  \hs{5} b \equiv -\frac{2f}{M}.
\label{ex:fermion_mix2}
\ee
This leads to the following eigenvalue equation
\be
m_{\rm F}^2 - (a+b)m_{\rm F} + ab - m_{{\rm B},2}^{2}=0,
\label{eq:ev2}
\ee
and so mass-eigenvalues are
\be
m_{{\rm F},1}^{} = \frac{a+b}{2} + 
\sqrt{ \left(\frac{a-b}{2} \right)^2 + m_{{\rm B},2}^2}\hs{3},  \hs{3}
m_{{\rm F},2}^{} =  \frac{a+b}{2} - 
\sqrt{ \left( \frac{a-b}{2} \right)^2 + m_{{\rm B},2}^2}\hs{3}.
\label{ex2_mass2}
\ee
By denoting corresponding mass-eigenstates as $\tilde{\psi}_1$
 and $\tilde{\psi}_2$, their mixing matrix becomes
\be
\left( \hspace{-1mm}
 \begin{array}{c} 
  \tau\\
  \lambda
 \end{array}\hspace{-1mm}
\right)
= \left( \hspace{-1mm} 
 \begin{array}{cc}
 V_{1,1} & \hs{-2} V_{1,2} \\
 V_{2,1} & \hs{-2} V_{2,2} \\
 \end{array}
\hspace{-1mm} \right)\hspace{-1mm} \left( \hspace{-1mm}
 \begin{array}{c} 
  \tilde{\psi}_1\\
  \tilde{\psi}_2
 \end{array}
\hspace{-1mm} \right), \hs{3}V_{1,1} 
= \frac{m_{{\rm F},1}^{}-b}{m_{{\rm B},2}^{}} V_{2,1}, 
\hs{3} V_{1,2} = \frac{m_{{\rm F},2}^{}-b}{m_{{\rm B},2}}^{} V_{2,2},
\ee
where we denote gauge boson and 
its mass as $(v_{i=2})_{\mu} \equiv (v_2)_{\mu}$ 
and $m_{{\rm B},2}^{}$ instead of $v_{\mu}$ and $m_{{\rm B}}^{}$.
Therefore NG fermion interaction Lagrangian Eq.(\ref{int_lag}) becomes
\bea
{\cal L}_{{\rm int}} \hs{-2}
&=&\hs{-2} \frac{V_{2,1}}{\sqrt{2}f} b\, 
\psng\,\sigma^{\mu\nu}\tilde{\psi}_1 (v_2)_{\mu\nu} 
- i \frac{V_{2,1}}{\sqrt{2}f}(-a m_{{\rm F},1}^{} +ab) 
\psng\, \sigma^\mu \, \bar{\tilde{\psi}}_1 \, (v_2)_{\mu} 
\nonumber \\
\hs{-2}&&\hs{-2}+ \frac{V_{2,2}}{\sqrt{2}f} b\, 
\psng\,\sigma^{\mu\nu}\tilde{\psi}_2 (v_2)_{\mu\nu} 
- i \frac{V_{2,2}}{\sqrt{2}f}(-a m_{{\rm F},2^{}} +ab) \psng\, 
\sigma^\mu \,\bar{\tilde{\psi}}_2\, (v_2)_{\mu} + {\rm h.c.},
\eea
and so these coupling constants are
\bea
\hef{2,1}^{(1)} \equiv \ds\frac{V_{2,1}}{\sqrt{2}f} b\,, &
\hef{2,1}^{(2)}  \equiv \ds - \frac{V_{2,1}}{\sqrt{2}f}
(-a m_{{\rm F},1}^{} +ab),\\
\hef{2,2}^{(1)}  \equiv \ds\frac{V_{2,2}}{\sqrt{2}f} b\,, &
\hef{2,2}^{(2)}  \equiv\ds - \frac{V_{2,2}}{\sqrt{2}f}
(-a m_{{\rm F},2}^{} +ab).
\eea

Now we consider the relation among these coupling and mass-splitting
\be
m_{{\rm B},2}^{2} - m_{{\rm F},i}^2 =  - (a+b)m_{{\rm F},i}^{} + ab.
\label{mass-splite} 
\ee
First we consider $m_{\rm B}^{} = 0$ and $m_{\rm F}^{} \ne 0 $.
 In this case, $m_{{\rm B},2}^{} = ek = 0$, 
 {\em i.e.}, $a = 0, \,b = m_{{\rm F},2}^{}$ 
 and there is no mixing $V_{2,2}=V_{1,1}=1,V_{1,2}=V_{2,1}=0$. 
 This result agrees with Eq.(\ref{v:GTR_1})
\be
\hef{2,2}^{(1)} = \frac{m_{{\rm F},2}^{}}{\sqrt{2}f}.
\ee
Next for $m_{\rm B}^{} \ne 0 $ and $m_{\rm F}^{} = 0 $, 
this is realized by taking $ab = m_{{\rm B},2}^{2} $ 
for $m_{{\rm F},2}^{}$ in Eq.(\ref{ex2_mass2}) and thus 
\be
\hef{2,2}^{(2)} = - \frac{V_{2,2}}{\sqrt{2}f}m_{{\rm B},2}^{2},
\ee
which is the result of Eq.(\ref{v:GTR_2}) with mixing. \\
For general cases of $m_{\rm B}^{} \ne 0 $ 
and $m_{\rm F}^{} \ne 0 $, 
the mass-splitting Eq.(\ref{mass-splite}) can be rewritten as 
\be
m_{{\rm B},2}^{2}  -m_{\rm F}^2   = -(a+b) m_{\rm F}^{} + ab 
= \left[ - a m_{\rm F}^{} + ab \,\right] +   \left[- m_{\rm F}^{}b \,\right] ,
\ee
and so the combination of two couplings $\hII + \,m_{\rm F} \hI$
leads
\bea
\left( \hII + \,m_{\rm F} \hI \right)_{2,1} 
\hs{-2}&=&\hs{-2} - \frac{V_{2,1}}{\sqrt{2}f} 
\,\left(m_{{\rm B},2}^{2}  -m_{{\rm F},1}^2 \right) ,\\
\left( \hII + \,m_{\rm F} \hI \right)_{2,2}
\hs{-2}&=&\hs{-2}  - \frac{V_{2,2}}{\sqrt{2}f} 
\,\left( m_{{\rm B},2}^{2}  -m_{{\rm F},2}^2  \right),
\eea
which agree with the SUSY Goldberger-Treiman relation 
with mixing in Eq.(\ref{v:GTR_12}). 

Thus the couplings and mass-splittings obtained 
in a SUSY breaking model discussed in this section 
all satisfy our results in Eqs.(\ref{v:GTR_1})-(\ref{v:GTR_12}).

%%%%%%%%%%%%%%%%%%%%%%%%%%%%%%%%%%%%%%%

\renewcommand{\thesubsection}{Acknowledgments}
\subsection{}

We thank Nobuhito Maru and Yutaka Sakamura for a collaboration 
where this work has started.

\end{document}